\documentclass[twocolumn]{aastex62}
\pdfoutput=1 %for arXiv submission
\usepackage{amsmath,amstext}
\usepackage[T1]{fontenc}
\usepackage{apjfonts} 
\usepackage[figure,figure*]{hypcap}
\usepackage{subfigure}
\usepackage{epsfig}
\usepackage{longtable}
\usepackage{isotope}
\newcommand{\cutt}[1]{\textcolor{blue}{}}

\newcommand{\Ms}{{\ensuremath{{M}_{\odot} }}}

\shorttitle{CSM / Binary interactions}
\shortauthors{Tsai et al.}

\begin{document}

\title{Interacting Binary Stars as Progenitors for Interacting Supernovae}

\author[0000-0001-5466-8274]{Sung-Han Tsai}
\affiliation{Department of Physics, National Taiwan University, Taipei 10617, Taiwan} 
\affiliation{Institute of Astronomy and Astrophysics, Academia Sinica, Taipei 106319, Taiwan} 
\email{m1069003@gm.astro.ncu.edu.tw}

\author[0000-0002-4848-5508]{Ke-Jung Chen}
\affiliation{Institute of Astronomy and Astrophysics, Academia Sinica, Taipei 106319, Taiwan} \affiliation{Heidelberger Institut für Theoretische Studien, Schloss-Wolfsbrunnenweg 35, 69118 Heidelberg, Germany}

\author[0000-0003-2611-7269]{Keiichi Maeda}
\affiliation{Department of Astronomy, Kyoto University, Kitashirakawa-Oiwake-cho, Sakyo-ku, Kyoto 606-8502, Japan}

\author[0000-0003-1295-8235]{Po-Sheng Ou}
\affiliation{Department of Physics, National Taiwan University, Taipei 10617, Taiwan} 
\affiliation{Institute of Astronomy and Astrophysics, Academia Sinica, Taipei 106319, Taiwan} 

\author[0000-0002-4460-0097]{Friedrich K.\ R{\"o}pke}
\affiliation{Zentrum für Astronomie der Universität Heidelberg, Astronomisches Rechenzentrum, M{\"o}nchhofstr.\ 12--14, 69120 Heidelberg, Germany}
\affiliation{Heidelberger Institut für Theoretische Studien, Schloss-Wolfsbrunnenweg 35, 69118 Heidelberg, Germany}
\affiliation{Zentrum für Astronomie der Universität Heidelberg, Institut für Theoretische Astrophysik, Philosophenweg 12, 69120 Heidelberg, Germany}

\begin{abstract}
Dense, compact circumstellar media (CSM) are required to power strongly interacting supernovae, yet their physical origin remains uncertain. We present a systematic study of binary stellar evolution models computed with \texttt{MESA}, demonstrating that Case C mass transfer—initiated after core helium ignition—naturally produces the dense, nearby CSM inferred in interacting events. Across a grid of binary models, we find that donors of $10$--$20\,M_\odot$ in binaries with separations of $\sim 1000$--$2700\,R_\odot$ undergo late-stage Roche-lobe overflow within $\sim 10^{3}$ yr prior to core collapse, ejecting $\sim 0.01$--$0.2\,M_\odot$ and forming CSM extending to $\sim 10^{16}$--$10^{18}\,$cm. Our results suggest that the Case~C mass transfer may account for $\sim 13\%$ of all core-collapse supernova (CCSN) progenitors, rather than representing a rare channel. A subset of these Case~C binaries produces CSM properties that are quantitatively in agreement with those inferred for interacting supernovae such as SN~2014C. In contrast to earlier binary interactions or single-star mass loss, Case C transfer operates at the right time and scale to shape the immediate pre-supernova environment without requiring ad hoc eruptive mechanisms. Our results identify late-stage binary interaction as a robust and physically motivated channel for producing the dense CSM that powers interacting supernovae.
\end{abstract}


\section{Introduction}

Interacting supernovae (Type IIn) represent a unique class of stellar explosions driven by the dynamic interaction between the ejected material from SNe and their surrounding circumstellar medium (CSM), which forms before the star's demise \citep{Chevalier1994, Chevalier2014, Nakaoka2018, Fraser2020}. This interaction can produce luminous emission across multiple wavelengths as kinetic energy from the ejecta is converted into radiation through shocks that collide with the dense CSM \citep[see][]{Chevalier1994, Chevalier2011}. Many of interacting supernovae show narrow emission lines in the early spectra, indicating the presence of slowly moving, dense material surrounding the progenitor star prior to explosion. These observational signatures suggest that a substantial amount of CSM may exist in the vicinity of massive stars at the time of core collapse.

However, the properties of the CSM inferred from observations are often difficult to reconcile with the typical stellar winds expected from single massive stars \citep{Smith2017, vanloon2025}. The densities and radial extents of the CSM surrounding some supernova progenitors suggest that their mass-loss histories may have been significantly enhanced in the final stages of stellar evolution \citep{Chevalier1994, Moriya2013, Smith2014_1}. Several mechanisms have been proposed to produce enhanced mass loss shortly before core collapse, including eruptive mass loss from luminous blue variables (LBVs), wave-driven mass loss triggered during late nuclear burning stages, and binary interactions \citep{Smith2006, Quataert2012, Shiode2014, Smith2014, Smith2017, ouchi2017,ouchi2019ApJ...877...92O,Ercolino2024}. Binary interaction is particularly promising, because most massive stars reside in binaries \citep{Sana2012, Langer2012, Ercolino2024, Marchant2024, vanloon2026}. Through mass transfer and envelope stripping, interacting binaries can significantly alter the mass-loss history and lead to the formation of dense CSM.

Mass transfer in interacting binaries is commonly classified into Cases~A, B, and C, depending on the evolutionary stage of the donor star \citep{paczy1971}. Case~A occurs when the donor fills its Roche lobe while still on the main sequence, burning hydrogen in its core. Case~B takes place after core hydrogen exhaustion but prior to helium ignition, during a phase of rapid stellar expansion. Case~C occurs after the onset of core helium burning, typically shortly before core collapse, and is therefore a promising channel for producing dense CSM in the immediate vicinity of the donor star prior to supernova. However, the properties of the CSM formed during these interactions—particularly in the final stages leading up to explosion—remain poorly understood.

To investigate how binary interactions shape the late-stage evolution of massive stars and their CSM, we perform detailed stellar evolution simulations using the \texttt{MESA} code. We model binary stars spanning a range of donor masses and orbital separations, identify the conditions under which strong mass transfer occurs, characterize the resulting CSM properties, and assess whether interacting binaries can serve as progenitors of interacting supernovae.

The structure of this paper is as follows. In Section~2, we describe the \texttt{MESA} stellar evolution code and the setup of our binary models. In Section~3, we present the binary evolution results and the physical properties of the resulting CSM. In Section~4, we discuss the astrophysical implications of our findings for interacting SNe. Finally, we summarize our conclusions in Section~5.

\section{Numerical Method}
We use the \texttt{MESA} code \citep[version r10108;][]{paxt11,paxt13,paxt15} to construct our stellar evolution models. \texttt{MESA} is a one-dimensional Lagrangian stellar evolution code that evolves stars under the assumption of hydrostatic equilibrium, and includes key microphysics such as nuclear reaction networks, convective mixing, and the equation of state, which are essential for modeling stellar evolution.

For binary models, we adopt solar metallicity ($Z = 0.02$). The initial donor mass spans $M_{1} = 10$--$40~M_{\odot}$ in steps of $2~M_{\odot}$, corresponding to the typical mass range of core collapse supernova (CCSN) progenitors. We consider two fixed initial mass ratios, $q_{2} \equiv M_{2}/M_{1} = 0.9$ and $0.6$, which set the companion mass $M_{2}$. For each binary model, we sample 41 initial orbital separations, $a = 500$--$4500~R_{\odot}$ in increments of $100~R_{\odot}$, covering both interacting and non-interacting regimes, motivated by observations of close massive binaries \citep{Moe2025}. In total, the grid comprises 656 binary models for each fixed mass ratio $q_{2}$.

\subsection{Model Setup}
In our stellar evolution calculations, we adopt the default parameters for massive stars from the \texttt{inlist\allowbreak\_massive\allowbreak\_default} setup in the \texttt{MESAstar} module. We use the \texttt{approx21 (Cr60)} nuclear reaction network, which includes 21 isotopes and accounts for $\alpha$-chain reactions, heavy-ion processes, hydrogen-burning cycles, photodisintegration of heavy nuclei, and energy losses from thermal neutrinos. Nuclear burning is solved self-consistently with the stellar structure equations under the assumption of hydrostatic equilibrium.

During and after helium burning, we apply Type II opacity tables \citep{Iglesias1996}. Convection is modeled using the Ledoux criterion, with Henyey mixing-length theory \citep{Henyey1965} and a mixing-length parameter of $\alpha = 1.5$. The choice of overshooting parameters remains uncertain and varies among different stellar evolution codes, with limited observational constraints. Here, we adopt a uniform overshooting efficiency of $f = 0.002$ at all convective boundaries, including both burning and non-burning regions as suggested by \citet{Ou2023,Ou2026}.

We use the \texttt{MESAbinary} module to coevolve the two stars in a binary system. Each binary is evolved until either the donor star forms a collapsing iron core or the system becomes a common-envelope candidate when mass transfer causes the companion star to overflow its Roche-lobe. In addition, some of our models are terminated due to numerical difficulties, such as strong mass transfer or unstable core burning, which lead to extremely small timesteps. All \texttt{MESA} files required to reproduce our results are available at \dataset[doi:10.5281/zenodo.20534639]
{https://doi.org/10.5281/zenodo.20534639}.

\subsection{Stellar Wind} \label{sec2.2}
Our stellar wind prescriptions are based on \citet{Ou2023}, which separates the stellar wind into three regimes: the hot wind \citep{Vink2000, Vink2001}, the cool wind \citep{deJager1988}, and the WR wind \citep{Nugis2000}. The key aspects of each prescription are summarized below.

For hot, massive OB stars ($T_{\mathrm{eff}} > 10^{4}\,\mathrm{K}$), we adopt the mass-loss prescription of \citet{Vink2000, Vink2001}, which provides metallicity-dependent wind mass-loss rates for radiatively driven stellar winds.

As stars evolve to cooler effective temperatures ($T_{\mathrm{eff}} < 10^{4}\,\mathrm{K}$), we switch to the empirical cool-wind prescription of \citet{deJager1988}, which is based on observationally inferred mass-loss rates of luminous cool stars. Following \citet{Ou2023}, we use a cool-wind prescription with an inexplicit metallicity scaling relation.

For Wolf--Rayet (WR) stars, we adopt the prescription of \citet{Nugis2000}, in which the mass-loss rate depends on the stellar luminosity, surface composition, and metallicity, including corrections for wind clumping.

These wind prescriptions are implemented within \texttt{MESA}'s built-in \texttt{Dutch} wind framework. Following \citet{Ou2023}, the hot-wind prescription completely dominates for $T_{\mathrm{eff}} > 1.2\times10^{4}\,\mathrm{K}$, while the cool-wind prescription dominates for $T_{\mathrm{eff}} < 8000\,\mathrm{K}$. Between $8000$ and $1.2\times10^{4}\,\mathrm{K}$, the mass-loss rate is determined by a linear interpolation between the hot and cold wind prescriptions. More details of the wind prescriptions are described in the Appendix of \citet{Ou2023}.

\subsection{Mass Transfer in Binary}

We consider two main channels for mass exchange: wind accretion and Roche-lobe overflow (RLOF). The wind mass transfer is based on the stellar wind prescriptions described in Section~\ref{sec2.2}. Both stars in the binary lose mass through stellar winds, a fraction of which can be accreted by stars.
The RLOF is treated following \citet{Tsai2023}, which adopts the Ritter scheme \citep{Ritter1988} to model the mass transfer between binary components when the donor star fills its Roche-lobe. 
The Roche-lobe radius of the donor star, $R_{\mathrm{RL,1}}$, can be approximated as \citep{Eggleton1983}:
\begin{equation}
\frac{R_{\mathrm{RL,1}}}{a} = 
\dfrac{0.49\,q_2^{-2/3}}{0.6\,q_2^{-2/3} + \ln(1 + q_2^{-1/3})},
\label{eq:roche_lobe}
\end{equation}
\vspace{2pt}
The mass transfer rate of the Ritter scheme is
\begin{equation}
\dot{M} = -{\dot{M}}_{0} \, \mathrm{exp} \left(\frac{R_{1} - R_{\mathrm{RL,1}}}{H_{\mathrm{\mathrm{P,1}}}/\gamma(q_{2})}\right),
\label{eq:xfer}
\end{equation}
where $R_{1}$ is the radii of the donor star, $H_{\mathrm{P,1}}$ is the pressure scale height at the photosphere of the donor star and
\begin{equation}
{\dot{M}}_{0} = \frac{2\pi}{\sqrt{e}} F_{1}(q_{2})\frac{R^{3}_{\mathrm{RL,1}}}{GM_{1}}\left(\frac{k_{\mathrm{B}}T_{\mathrm{eff}}}{m_{\mathrm{p}}\mu_{\mathrm{ph}}}\right)^{3/2}\rho_{\mathrm{ph}},
\end{equation}
where $e$ is Euler's number, $k_{\mathrm{B}}$ is the Boltzmann constant, $m_{\mathrm{p}}$ is the proton mass, $T_{\mathrm{eff}}$ is the effective temperature of the donor, $\mu_{\mathrm{ph}}$ and $\rho_{\mathrm{ph}}$ are the mean molecular weight and density in its photosphere, and $F_1(q_2)$ and $\gamma(q_2)$ are fitting functions \citep[equations 15 and 16 in][]{paxt15}.

In this work, we assume a mass transfer efficiency of $50\%$ for both wind accretion and RLOF; i.e., the companion accretes half of the transferred mass, while the remaining half is ejected from the binary, resulting in CSM.

\subsection{CSM Formation}
We construct the CSM profile using the standard steady-state wind density relation,
\begin{equation}\label{eq:csmdes}
\rho_{\rm CSM}(r) = \frac{\dot{M}}{4\pi r^{2}v},
\end{equation}
where $\rho_{\rm CSM}$ is the density at radius $r$, $\dot{M}$ is the mass-loss rate, and $v$ is the wind or outflow velocity.

In our binary models, $\dot{M}$ includes contributions from both stellar winds and RLOF. Although Equation~\ref{eq:csmdes} assumes steady-state outflow, we apply it locally at each timestep of the \texttt{MESA} evolution. Using time-dependent mass-loss rates and surface escape velocities, we reconstruct a CSM density profile that captures the mass-loss history. This approach accounts for enhanced, episodic mass loss during binary interactions, which can differ significantly from the steady wind-driven CSM in single stars.

\section{Results}
\subsection{Possible Fates of $10-40\,\Ms$ Stars} \label{sec3.1}

\begin{figure}
\centering
\includegraphics[width=0.45\textwidth]{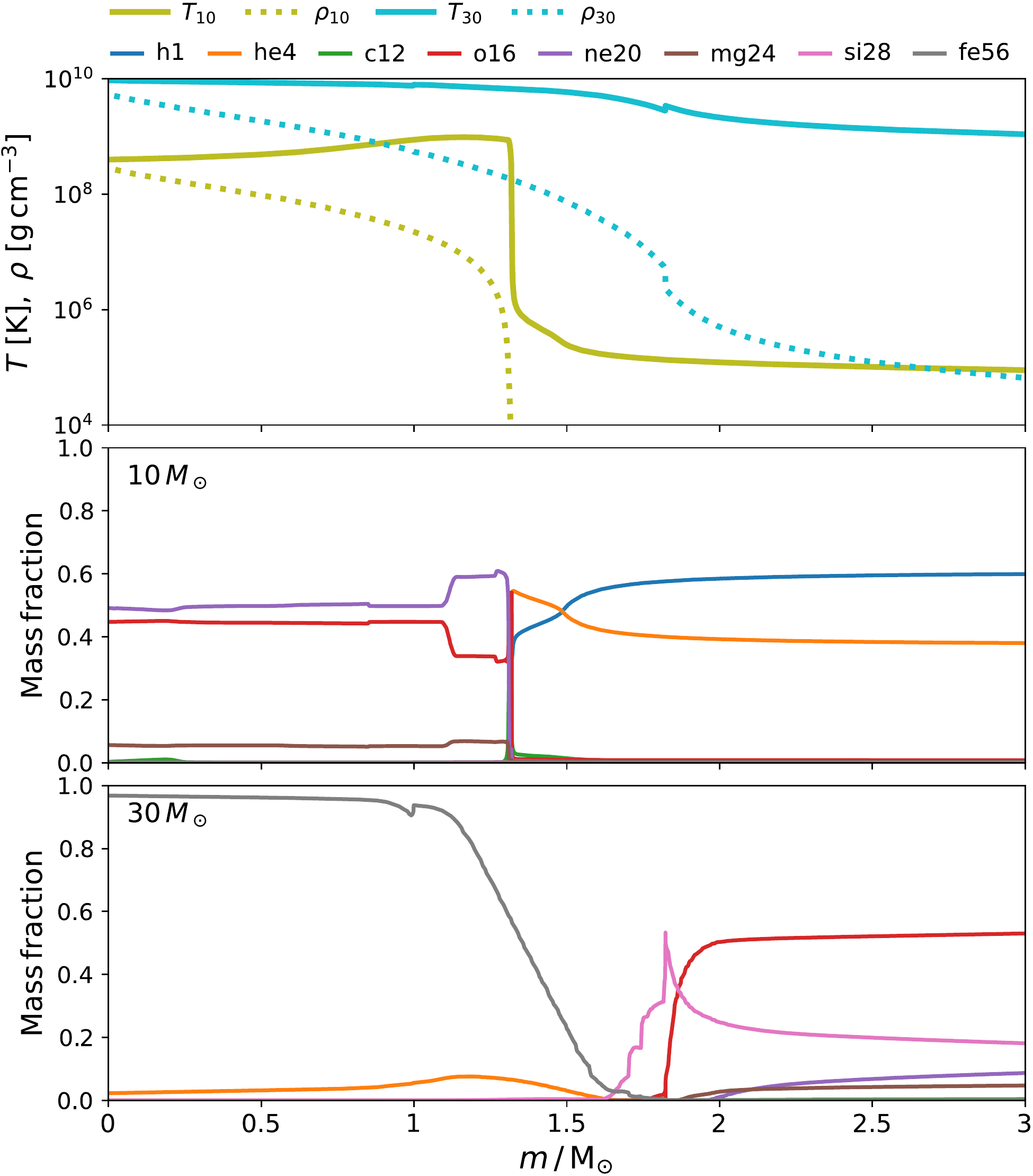}
\label{abundance}
\caption{Density, temperature, and elemental abundance profiles for the $10~M_{\odot}$ and $30~M_{\odot}$ stars at the end of their evolution. The top panel shows temperature ($T$; solid lines) and density ($\rho$; dotted lines) for both models. The middle and bottom panels present the corresponding abundance profiles. The $10~M_{\odot}$ star develops a degenerate oxygen–neon–magnesium (O–Ne–Mg) core, while the $30~M_{\odot}$ model forms an iron core prior to collapse.}

\end{figure}

To assess the evolutionary outcomes of binary stars, we first examine single-star models with initial masses of $10$--$40\,\Ms$ for the donor stars.  As shown in Figure~\ref{abundance}, the $10~M_{\odot}$ model develops a $1.329~M_{\odot}$ oxygen--neon--magnesium (O--Ne--Mg) core, but its central temperature and density remain insufficient to ignite further nuclear burning before strong electron degeneracy sets in. As the core mass approaches the Chandrasekhar limit ($\sim 1.375\,M_{\odot}$; \citealt{Nomoto1987}), electron captures are expected to trigger rapid contraction, potentially leading to an electron-capture supernova(ECSN; \citealt{Nomoto1984,Nomoto1987,Takahashi2013}). We note, however, that alternative thermonuclear outcomes have also been proposed for such progenitors \citep{holas2024}. In this work, however, we simply assume that $\sim 10~M_{\odot}$ stars explode through the collapse of an O–Ne–Mg core.

Due to numerical limitations, the model cannot be evolved until its final fate. We therefore estimate the remaining evolutionary timescale using the late-time growth of the O--Ne--Mg core. Specifically, we fit the final phase of core growth with a fifth-degree polynomial and extrapolate the time required for the core to reach the critical mass. This procedure yields a remaining lifetime of approximately $4\times10^{3}$ yr. We emphasize that this value should be regarded as an approximate estimate rather than a precise prediction, since late-stage shell burning episodes or structural readjustments may alter the actual growth rate of the O--Ne--Mg core.

The stars in the $12$ and $14~M_{\odot}$ models form an iron core during the silicon-burning stage, but terminate due to small timestep issues, while stars with initial masses of $16$–$40~M_{\odot}$ proceed to form dynamically collapsing iron cores. Therefore, we classify $12-40~M_{\odot}$ stars CCSN progenitors from a collapsing iron core.
As illustrated by the $30~M_{\odot}$ model in Figure~\ref{abundance}, the star develops an iron core approaching the Chandrasekhar mass and is expected to explode as a CCSN.

% Figure 1

\subsection{Binary Evolution Models}
We present the distribution of binary interaction types as a function of donor mass and orbital separation for a mass ratio of $q_{2} = 0.9$ \footnote{For comparison, the results for $q_{2} = 0.6$ are presented in Appendix A.} in Figure~\ref{mt_class_q09}. Among the 656 models, 70.6\% undergo mass transfer, including 64.2\% Case~B and 6.4\% Case~C, with four binaries experiencing both phases. Our results suggest that $70.6\%$ massive binaries undergo interactions. 
We note that the fractions here are based simply on the model number count and that the intrinsic frequency should be computed by taking into account the initial mass function and the distribution of the initial separations. Nevertheless, these numbers demonstrate that the binary interaction is quite common, in good agreement with the observations of \citet{Sana2012} (See Section~\ref{sec4.4} for details). Case~C mass transfer occurs only for donor masses $\lesssim 20~M_{\odot}$, as these stars undergo significant radial expansion during core carbon burning. For the $10\,M_{\odot}$ models, the star is expected to end its lives as a CCSN through the ECSN channel, as discussed in Section~\ref{sec3.1}. The exact remaining lifetime and the duration of the Case~C mass-transfer phase are uncertain because the calculations encountered numerical difficulties before reaching core collapse. Nevertheless, the RLOF had already commenced prior to the termination of the simulations, and therefore the classification of these systems as Case~C binaries remains robust. 

For donor masses $\lesssim 30~M_{\odot}$ and orbital separations $< 2500\,R_\odot$, more than 50\% of the models enter common-envelope evolution (CEE). This is mainly driven by rapid mass accretion onto the companion during RLOF, causing it to expand and potentially overfill its own Roche lobe, leading to unstable mass transfer during Case~B. In contrast, for donor masses $M_{1} > 30~M_{\odot}$, the companion’s response to mass accretion is more moderate: its envelope expands slightly and then gradually contracts during the mass transfer phase, preventing runaway expansion. As a result, these binaries can maintain stable Case~B mass transfer.

% Figure 2

\begin{figure*}
\centering
\includegraphics[width=1.\textwidth]{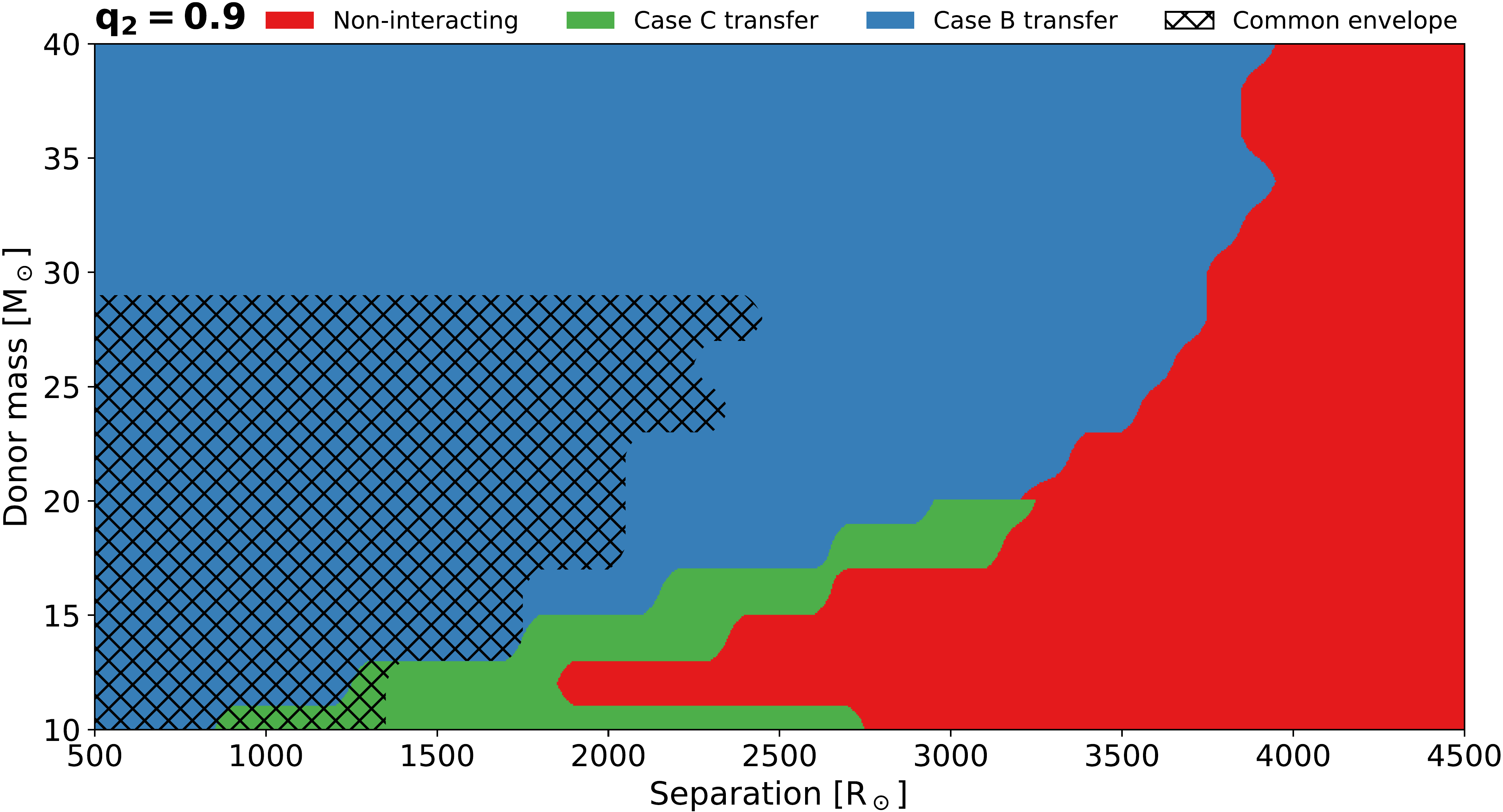}
\label{mt_class_q09}
\caption{Binary interaction outcomes—non-interacting (red), Case~C transfer (green), and Case~B transfer (blue)—as a function of initial orbital separation and donor mass $M_{1}$, assuming a fixed mass ratio $q_{2}$ of 0.9. Hatched regions indicate models that undergo common-envelope evolution (CEE). The boundary between interacting and non-interacting binaries shifts to larger separations with increasing total binary mass. A complex interaction landscape emerges at $M_{1} \lesssim 20,M_{\odot}$ and separations $\lesssim 3200~R_{\odot}$, where multiple interaction channels coexist. CEE occurs predominantly for $M_{1} \lesssim 28~M_{\odot}$ and separations $\lesssim 2500~R_{\odot}$.}
\end{figure*}

\subsection{Mass Transfer Rates}

Stellar winds and RLOF not only influence stellar evolution but also shape the CSM surrounding the binary system. Therefore, investigating the mass-loss histories of interacting binaries is essential to understand the resulting CSM structures.

\begin{figure*}
\centering
\includegraphics[width=1.\textwidth]{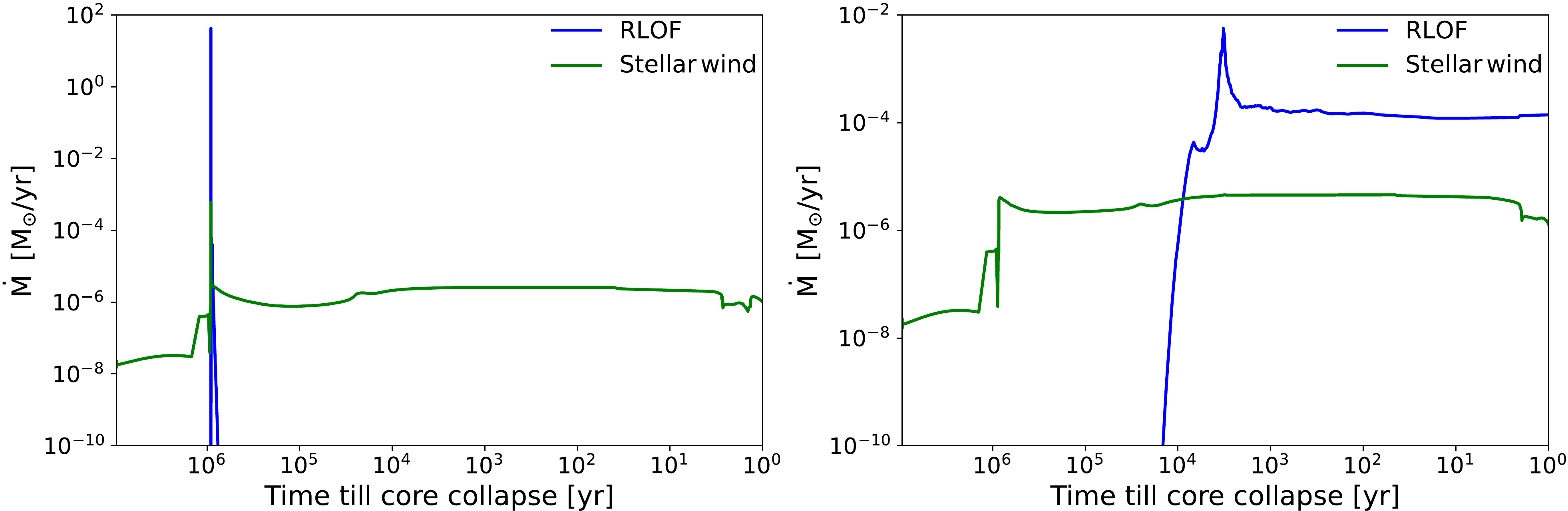}
\label{mdot}
\caption{Mass-loss history of an $18~M_{\odot}$ donor star undergoing binary interaction. The left panel shows the Case~B model, and the right panel shows the Case~C model. The companion mass is $16.2~M_{\odot}$, with initial orbital separations of $2400~R_{\odot}$ (Case~B) and $3000~R_{\odot}$ (Case~C). The x-axis indicates time before core collapse, and the y-axis shows the mass-loss rate. Although Case~B can reach high mass-loss rates, this occurs $\sim 1$ Myr before core collapse and therefore does not contribute to the formation of dense CSM immediately prior to explosion. In contrast, Case~C mass transfer occurs $\sim 10^{4}$ yr before core collapse and can persist until explosion.}
\end{figure*}

Figure~\ref{mdot} shows the mass-loss histories of the interacting binaries, one experiencing a Case~B transfer and the other Case~C transfer. These binaries initially have a $18\,M_{\odot}$ donor and a $16.2\,M_{\odot}$ companion, but with different initial orbital separations; $2400\,R_{\odot}$ for Case~B and $3000\,R_{\odot}$ for Case~C. The Mass-loss history from stellar wind for two models follows nearly identical evolutionary trends. During the main-sequence phase, the wind is dominated by hot-star winds. As the star evolves into an early-type supergiant, it experiences the bi-stability jump \citep{Pauldrach1990}. Eventually, the wind transitions from hot winds to cool supergiant winds, as described by \citet{Ou2023}.

The RLOF phase has little impact on the stellar wind evolution because it begins only after the wind has already transitioned from hot to cool supergiant winds. Furthermore, in both cases, the RLOF phase fails to completely remove the hydrogen envelope of the donor star, allowing stellar winds to continue during the post-main-sequence evolution with typical mass-loss rates of $\sim 
10^{-6}~M_{\odot}\,{\rm yr^{-1}}$.
In contrast, the RLOF mass-loss histories differ significantly between Case~B and Case~C binaries. In Case~B binaries, the mass transfer rate increases from $\sim10^{-5}$ to $\sim10^{-2}\,M_{\odot}\,{\rm yr^{-1}}$ within the first $\sim600$~yr after the onset of RLOF. Then it increases rapidly to reach a brief peak of $\sim10\,M_{\odot}\,{\rm yr^{-1}}$ for $\ll 1$~yr. The subsequent mass transfer maintains $\sim10^{-5}\,M_{\odot}\,{\rm yr^{-1}}$ for $\sim4.3\times10^{4}$~yr. Because Case~B mass transfer occurs soon after the star leaves the main sequence, it is generally too early to produce the dense CSM required for interacting supernovae.

For Case~C binaries, RLOF dominates the mass loss during the final $\sim 10^{4}$ yr prior to core collapse, reaching peak rates of $\sim 10^{-3}~~M_{\odot}\,{\rm yr^{-1}}$. This late-stage interaction sustains mass transfer at $\sim 10^{-4}~~M_{\odot}\,{\rm yr^{-1}}$ up to the time of explosion.

We calculate the total mass lost via RLOF ($\Delta M_{\rm RLOF}$) and stellar winds ($\Delta M_{\rm wind}$) during the final $1000$~yr before core collapse. In the Case~B model, RLOF does not contribute during this phase; mass loss is instead dominated by stellar winds, with $\Delta M_{\rm wind} = 2.57 \times 10^{-3}~M_{\odot}$. In contrast, the Case~C model exhibits a comparable wind contribution ($\Delta M_{\rm wind} = 4.52 \times 10^{-3}~M_{\odot}$), but the mass loss is dominated by RLOF, which contributes $\Delta M_{\rm RLOF} = 1.60 \times 10^{-1}~M_{\odot}$. This results in the formation of a denser CSM surrounding the donor star compared to that produced by stellar winds alone. Thus, the Case~B and Case~C binaries produce distinct CSM structures due to the different timings of the onset of mass transfer, as explored in the following section.

% Figure 3

\subsection{Physical Properties of CSM} \label{sec:csm_properties}

\begin{figure*}
\centering
\includegraphics[width=1.\textwidth]{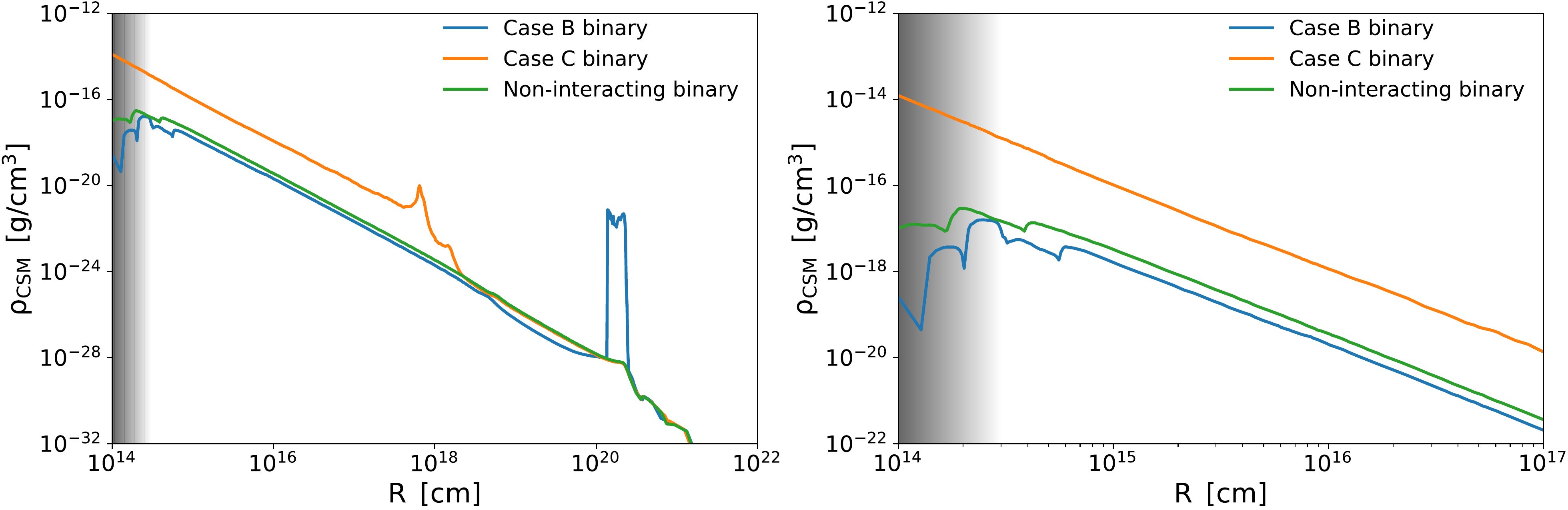}
\label{csm_density}
\caption{CSM density profiles for a binary system with an $18~M_{\odot}$ donor and a $16.2~M_{\odot}$ companion, constructed from the mass-loss history shown in Figure~\ref{mdot}. The left panel shows the full radial extent of the CSM, while the right panel zooms in on the inner region relevant to interacting supernovae. The gray shaded region at $r \lesssim 3 \times 10^{14}\,\mathrm{cm}$ indicates the approximate binary orbital separation; regions within this radius are excluded from the CSM profile. Compared to the non-interacting case, binary interactions produce denser and more structured CSM, including shell-like features during RLOF.}
\end{figure*}

Figure~\ref{csm_density} shows the CSM density profiles for binary models initially with an $18~M_{\odot}$ donor and a $16.2~M_{\odot}$ companion with different orbital separations, covering Case~B, Case~C and non-interacting binaries. The resulting density profiles broadly follow the $\rho \propto r^{-2}$ scaling expected for steady winds, but binary interactions introduce distinct deviations from this profile.

For the Case~B model, a density enhancement appears at $r \sim 10^{20}$ cm from the binary center of mass. This feature corresponds to the RLOF phase that occurs as the donor evolves off the main sequence. During this phase, part of the hydrogen envelope is stripped from the donor, reducing subsequent wind mass loss compared to the non-interacting case. As a result, the CSM density in the Case~B model is lower than in the non-interacting system over the radial range $10^{14}$–$10^{20}$ cm.

For the Case~C model, RLOF occurs during the donor’s second radial expansion in core carbon burning, approximately $\sim 10^{4}$ yr before core collapse. The resulting increase in the rate of mass-loss produces a prominent enhancement of density in the CSM at $r \sim 10^{14}$–$10^{18}$ cm as shown in Figure~\ref{csm_density}.
Because RLOF dominates the mass loss up to core collapse, with a characteristic rate of $\sim 10^{-3}~M_{\odot}\,{\rm yr^{-1}}$ about two orders of magnitude higher than stellar winds, the CSM is significantly denser than in both the Case~B and non-interacting models.

The innermost CSM structure (which might leave observational signatures in the first few days after the SN) is also affected. 
For the Case~B and non-interacting binaries, small fluctuations at radial scales of $\sim10^{14}$~cm originate from structural instabilities during the final $\sim10$~yr of stellar evolution, which cause variations in the mass-loss rate by the stellar wind. This density fluctuation is smoothed out in Case~C since the RLOF dominates the mass loss. 

These results demonstrate that binary interactions can significantly modify the CSM density structure. Different mass transfer channels leave characteristic marks on the density profile across multiple radial scales, providing potential observational signatures that link pre-supernova mass loss to the binary evolution history of the progenitor system.

% Figure 4

\section{Discussion}
\subsection{Physics of Case~C Transfer}

Table~\ref{tab:caseC_models} summarizes the physical characteristics of the Case~C mass transfer in models that evolve to the collapse of the core. For donor masses of $14$ and $18~M_{\odot}$, the onset and duration of Case~C interaction depend sensitively on the initial orbital separation. Binaries with smaller separations initiate mass transfer earlier, but the interaction is less likely to persist until the core collapse. As a result, the contribution of RLOF to mass loss during the final $\sim 10^{3}$~year is reduced. In contrast, wider binaries tend to initiate the Case~C interaction later and maintain the RLOF until the core collapse, resulting in larger $\Delta M_{\rm RLOF}$ in the final stages of evolution, as shown in Table~\ref{tab:caseC_models}. Consequently, mass loss in the final $\sim 10^{3}$~year is dominated by the wind in the closer binaries and by RLOF in the wider binaries. The Case~C mass transfer for a donor $20~M_{\odot}$ is considerably weaker than in the $14$ and $18~M_{\odot}$ models. The $20~M_{\odot}$ star undergoes a less pronounced secondary expansion following core helium ignition, resulting in only brief or marginal RLOF episodes. Consequently, the RLOF phase terminates well before core collapse, and the mass transfer rate subsequently declines to $\sim10^{-9}~M_{\odot}\,{\rm yr^{-1}}$, two orders of magnitude below the values found in the lower-mass Case~C systems. As a result, the final $\sim10^{3}$ yr of evolution are almost entirely dominated by stellar winds rather than RLOF.

Despite the limited duration of RLOF in many cases, some models experience substantial mass removal, with $\Delta M_{\rm RLOF} \sim 10^{-2}$–$10^{-1}~M_{\odot}$. This indicates that Case~C mass transfer can contribute considerable to the overall mass-loss budget. These results highlight that the key characteristics of Case~C transfer are governed by the interplay between stellar expansion, orbital separation, and the evolving structure of the donor star, rather than by a steady or quasi-steady mass transfer process.

\begin{table*}[ht]
\centering
\begin{tabular}{cccccc}
\hline
\hline
$M_1$ [$M_\odot$] & $a$ [$R_\odot$] & Case~C duration [yr] & $\Delta M_{\rm RLOF}$ [$M_\odot$] & $\Delta M_{\rm wind}$ [$M_\odot$] & Case~C until CC? \\
\hline

14 & 1800 & $2.60 \times 10^{4}$ & $7.76 \times 10^{-4}$ & $1.53 \times 10^{-3}$ & No \\
   & 1900 & $2.38 \times 10^{4}$ & $2.69 \times 10^{-4}$ & $1.41 \times 10^{-3}$ & No \\
   & 2200 & $3.98 \times 10^{3}$ & $2.63 \times 10^{-1}$ & $1.65 \times 10^{-3}$ & Yes \\
   & 2300 & $1.63 \times 10^{3}$ & $5.67 \times 10^{-3}$ & $1.87 \times 10^{-3}$ & Yes \\
\hline

18 & 2700 & $2.15 \times 10^{4}$ & $8.33 \times 10^{-4}$ & $4.45 \times 10^{-3}$ & No \\
   & 2900 & $1.13 \times 10^{4}$ & $2.68 \times 10^{-2}$ & $4.28 \times 10^{-3}$ & Yes \\
   & 3000 & $8.60 \times 10^{3}$ & $1.60 \times 10^{-1}$ & $4.52 \times 10^{-3}$ & Yes \\
   & 3100 & $3.48 \times 10^{3}$ & $6.66 \times 10^{-2}$ & $4.90 \times 10^{-3}$ & Yes \\
\hline

20 & 3000 & $1.49 \times 10^{4}$ & $4.67 \times 10^{-5}$ & $6.56 \times 10^{-3}$ & No \\
   & 3100 & $4.61 \times 10^{3}$ & $7.26 \times 10^{-6}$ & $6.34 \times 10^{-3}$ & No \\
   & 3200 & $7.46 \times 10^{2}$ & $2.29 \times 10^{-7}$ & $6.37 \times 10^{-3}$ & No \\

\hline
\end{tabular}

\caption{
Summary of the Case~C binary models and their mass-loss properties. The table lists the initial donor mass ($M_1$), initial orbital separation ($a$), duration of the Case~C mass transfer phase, mass lost through RLOF ($\Delta M_{\rm RLOF}$) and stellar winds 
($\Delta M_{\rm wind}$) {\it during the final $1000$~yr prior to core collapse}. The Case~C duration is defined as the period during which the mass-loss rate from RLOF exceeds that from stellar winds, beginning when RLOF first dominates and ending when the RLOF rate falls below the wind mass-loss rate or at core collapse. The column "Case~C until CC?" indicates whether the Case~C mass transfer persists until the core collapse of the donor star. The table shows that only binaries with donor masses up to $\sim18~M_\odot$ sustain Case~C mass transfer until core collapse.}

\label{tab:caseC_models}
\end{table*}

\subsection{Formation of Dense CSM}

Recent observations of interacting hydrogen-rich supernovae, including Type~IIn and rapidly evolving Type~II events, suggest that dense and structured CSM is common among massive-star explosions \citep[e.g.,][]{Charalampopoulos2025, Goto2025, Reyonlds2025, Hillenkamp2026}. Such systems require enhanced pre-supernova mass loss, sometimes involving both dense inner CSM and more extended material. These characteristics are consistent with the time-dependent mass-loss histories produced by Case~C interaction \citep[see also][]{ouchi2017}. 

To estimate the radial extent of the CSM, we adopt the surface escape velocity from the \texttt{MESA} models as a proxy for the outflow velocity of $\sim 10^{6}~\rm cm\,s^{-1}$ across our models. Combined with the duration of the Case~C interaction of $\sim10^{3}$ -- $2.6\times10^{4}$~yr based on our models, the ejected material can reach distances of $\sim10^{16}$ -- $10^{18}$~cm before core collapse. Given that the characteristic size of the binary is $\sim10^{14}$--$10^{15}$~cm, this implies that the inner region at 
$r\sim10^{15}$--$10^{16}$~cm is also part of the dense CSM regime. Therefore, Case~C mass transfer can naturally produce dense CSM on a wide range of radial scale. Specifically, it can produce either a dense CSM extending to $\sim 10^{16}$--$10^{18}$~cm, potentially corresponding to SNe~IIn, or a dense CSM shell at similar radii separated from the progenitor by a low-density cavity, potentially corresponding to rebrightening supernovae (see Section~\ref{sec4.3} for further details).

Another question is whether this scenario can also account for the dense and confined CSM frequently inferred around SNe~IIP/L; for example, the dense and spatially compact CSM inferred for iPTF13dqy \citep{yaron2017}, as well as the close-in CSM revealed by early-time spectroscopy of SN~2023ixf \citep{JacobsonGalan2023}, imply material located at smaller radii than the most extended CSM predicted by our models. For SN~2023ixf, early-time spectra indicate dense, close-in CSM photoionized by SN radiation, with spectral evolution suggesting a decline in density beyond $r\gtrsim10^{15}$~cm \citep{JacobsonGalan2023}. The discrepancy in radial scale does not necessarily rule out a Case~C origin. Our reconstruction assumes a characteristic outflow velocity and spherical expansion. In reality, mass lost through RLOF is a multidimensional process, unlikely to be isotropic. Instead of a symmetric outflow, transferred material may accumulate in the orbital plane, leading to equatorial outflows or circumbinary structures shaped by binary motion. Here, the same mass occupies a smaller solid angle, causing higher local densities than those from spherical averaging.

The combination of anisotropic mass loss and reduced expansion velocities could keep dense material at smaller radii for longer periods, potentially answering the tension between our reconstructed CSM profiles and the compact CSM inferred for objects such as SN~2023ixf.

Our results indicate that the Case~C binary interaction naturally produces dense CSM, while the degree of spatial confinement depends sensitively on the outflow velocity and geometry. Multi-dimensional hydrodynamic simulations are therefore required to determine the final distribution of the CSM formed through binary interaction.

\subsection{Implications on Diversity of Interacting Supernovae} \label{sec4.3}

Binary interaction provides a natural framework for understanding the diversity of interacting supernovae through its impact on both the stellar envelope and the CSM. Based on our binary models, the evolution can be broadly categorized into four regimes: Case~B mass transfer with and without CEE, and Case~C mass transfer with and without CEE. These regimes lead to distinct combinations of hydrogen-envelope maintenance and CSM structure.

Case~B mass transfer efficiently removes the hydrogen envelope of the donor star, particularly when accompanied by CEE. As a result, these models naturally produce stripped-envelope progenitors, leading to Type~IIb or Type~Ib/Ic supernovae \citep{yoon2017ApJ...840...10Y,ouchi2017}; the initial separation can also lead to diverse properties of SNe IIb \citep{maeda2015ApJ...807...35M}. In contrast, Case~C mass transfer typically occurs shortly before core collapse and removes only a fraction of the hydrogen envelope, leaving the final supernova type 
largely unchanged (i.e., Type~IIP/IIL). To connect the binary evolution models with the observable supernova properties, we classify the supernova type based on the remaining hydrogen-envelope mass of the progenitor. Following \citet{ouchi2017}, models with $M_{\rm H} > 1\,M_{\odot}$ are classified as Type~IIP/IIL supernovae, those with $M_{\rm H} < 0.01\,M_{\odot}$ as Type~Ib/Ic supernovae, and models with $0.01\,M_{\odot} < M_{\rm H} < 1\,M_{\odot}$ as Type~IIb progenitors. Representative examples of these evolutionary outcomes are summarized in Table~\ref{tab:binary_18_36}, which lists the final stellar properties and predicted supernova types for selected binary models. Table~\ref{tab:binary_18_36} illustrates how different interaction channels, particularly Case~B mass transfer and its stability, lead to different hydrogen-envelope masses and the corresponding SN classifications. However, the Case~C interaction can significantly modify the surrounding CSM, especially when mass transfer occurs within $\sim10^{3}$–$10^{4}$~yr before explosion; in this case, the SN may be even categorized to type IIn. 

A key result of our models is that the Case~C interaction can bifurcate into distinct evolutionary pathways depending on the binary separation. In our grid, this bifurcation is most clearly seen for lower-mass donors. For example, for $10$--$12~M_{\odot}$ donor stars, binaries with separations $\lesssim 1300~R_{\odot}$ consistently enter a common-envelope phase following the onset of Case~C mass transfer, while binary models with slightly larger separations remain in stable mass transfer without developing a common-envelope. Because the donor’s radial expansion timescale in this phase is short, the outcome of Case~C mass transfer is highly sensitive to the orbital separation; even small differences can lead to a bifurcation between stable and unstable evolution.

This bifurcation provides an explanation for the diversity observed in some interacting SNe, specifically those showing "rebrightening" in radio and X-rays. SN Ib 2014C, which exhibited a transition from a hydrogen-free type~Ib SN to a strongly interacting event, has been interpreted to originate from a system surrounded by a hydrogen-rich dense CSM shell at $\sim10^{16}$–$10^{17}$~cm \citep{margutti2017}. In our framework, such a configuration can arise from Case~C mass transfer followed by CEE, where the envelope is rapidly ejected during the spiral-in of the companion \citep{ivnova2013, Kruckow2016, Wei2024}, forming a detached, hydrogen-rich CSM shell plus a hydrogen-free SN Ib progenitor.
Hydrodynamic simulations further suggest that the enhanced mass-loss phase responsible for the dense shell occurred $\sim 5000-1000$~yr before explosion, with mass-loss rates reaching $\sim8\times10^{-4}~M_{\odot}~{\rm yr^{-1}}$ \citep{orlando2024}. These timescales and mass-loss rates are broadly consistent with those produced by Case~C interaction in our models. However, the hydrogen-poor Type~Ib classification of SN~2014C indicates that most of the hydrogen envelope must have been removed before explosion, whereas stable Case~C mass transfer in our models generally retains a substantial hydrogen envelope and therefore produces Type~II progenitors. In addition, the presence of a low-density cavity interior to the dense shell suggests that the enhanced mass-loss episode terminated before core collapse. We therefore suggest a scenario in which Case~C interaction becomes unstable and triggers a common-envelope episode. Following the envelope ejection, subsequent winds from the compact progenitor can clear the inner region and generate the cavity inferred by \citet{orlando2024}. Furthermore, \citet{orlando2024} proposed a Case~C mass transfer scenario followed by a common-envelope episode for SN~2014C by adopting a backward approach, reconstructing the pre-explosion CSM from the observed X-ray data. In contrast, our work follows the forward evolution of binary stars from the zero-age main sequence. Despite these fundamentally different methodologies, both studies converge on a common-envelope origin for the dense CSM surrounding SN~2014C. Thus, this progenitor scenario of SN~2014C is independently supported by two complementary approaches, providing a self-consistent explanation for the observed CSM structure.

On the other hand, hydrogen-rich SN II 2018ivc shows evidence for a less extreme situation -- the relatively dense CSM shell at $\leq 10^{17}$ cm with inferred mass-loss rates of $\sim10^{-3}~M_{\odot}\,{\rm yr^{-1}}$ \citep{maeda2023}. In contrast to the case for SN 2014C, the inner CSM is also substantially dense, corresponding to the mass-loss rate of $\sim10^{-4}~M_{\odot}\,{\rm yr^{-1}}$ \citep{maeda2023ApJ...942...17M}. Namely, the mass-loss history has been inferred to show a sudden increase to $\sim10^{-3}~M_{\odot}\,{\rm yr^{-1}}$ at $\sim 1000$ yr before the explosion, followed by the moderately high rate of $\sim10^{-4}~M_{\odot}\,{\rm yr^{-1}}$ toward the SN. This behavior is explained by our model, as a system experiencing a stable Case~C mass transfer that starts to operate very close ($<1000$ yr) to the core collapse (e.g., Figure~\ref{mdot}). We therefore suggest a possible unification of the progenitor models of the "rebrightening" SNe, as demonstrated by SNe 2014C and 2018ivc \citep[see also][]{maeda2023} -- these might indeed be similar binary models, while a small difference in the separation leads to the bifurcation; an SN Ib surrounded by a very dense CSM shell with a cavity, and an SN II surrounded by a moderately dense CSM shell but without a rarefied cavity. Depending on the timing of the initiation of the Case C interaction (controlled by the initial separation), the scenario might be further linked to some SNe IIn. 

An important implication of this interpretation is that these rebrightening or strongly interacting SNe may originate from a binary system with relatively lower mass donor stars that undergo Case~C interaction. This contrasts with alternative scenarios involving more massive stars -- LBV (luminous blue variable)-like progenitors for hydrogen-rich cases \citep{nagao2026A&A...708A.294N}, and Wolf--Rayet (WR) progenitors for hydrogen-poor/free cases such as SNe~Ibn/Icn \citep{Pastorello2007,maeda2022ApJ...927...25M} or SNe~Ic-CSM \citep{Maeda2026}.

Our results suggest that the diversity of interacting SNe may be understood within a unified binary-interaction framework, where the timing and stability of Case~C mass transfer, particularly whether the system enters a CEE, plays a central role in shaping both the progenitor structure and the surrounding CSM.

\begin{table*}[ht]
\centering
\begin{tabular}{ccccccc}
\hline
\hline
$M_1$ [$M_\odot$] & $a$ [$R_{\odot}$] & Lifetime [Myr] & $M_{f}$ [$M_{\odot}$] & H-layer [$M_{\odot}$] & RLOF Type & Predicted SN Type \\
\hline
18 & 2000  & 8.63 & 10.34 & 6.09 & Case~B & C.E. \\
   & 2400  & 9.54 & 6.04 & 0.54 & Case~B & CCSN, IIb \\
   & 2700  & 9.55 & 9.47 & 3.91 & Case~C & CCSN, IIP/IIL (IIn)\\
   & 3200  & 9.50 & 14.34 & 8.86 & non-interacting & CCSN, IIP/IIL \\
\hline
36 & 3000  & 5.07 & 11.89 & 0.00 & Case~B & CCSN, Ib/Ic \\
   & 3600  & 5.04 & 14.33 & 0.50 & Case~B & CCSN, IIb \\
   & 3800  & 5.07 & 15.21 & 1.33 & Case~B & CCSN, IIP/IIL \\
   & 4000  & 5.06 & 15.24 & 1.38 & non-interacting & CCSN, IIP/IIL \\
\hline
\end{tabular}
\caption{Summary of binary models with $18~M_{\odot}$ and $36~M_{\odot}$ donors. Each model is evolved until either core collapse or the onset of CEE phase. The table lists the stellar lifetime, final mass ($M_{f}$), remaining hydrogen-envelope mass, the corresponding RLOF and predicted supernova types. Depending on the initial orbital separation, the binaries undergo different interaction channels—non-interacting, Case~B, or Case~C—which significantly influence the evolution of the donor star and its resulting SN type.}
\label{tab:binary_18_36}
\end{table*}

\subsection{Contribution of Case~C Mass Transfer to Core Collapse Supernovae} \label{sec4.4}

We now estimate the rate of occurrence of Case~C mass transfer based on our results with an initial mass function (IMF) and an assumed orbital-separation distribution. For the IMF, we adopt the Kroupa IMF \citep{Kroupa2001,Kroupa2002}. In the massive star regime relevant to this work, this can be approximated as
\begin{equation}
    \xi(M_1) \equiv \frac{dN}{dM_1} \propto M_1^{-\alpha},
    \qquad \alpha = 2.3 ,
\end{equation}
where \(M_1\) is the donor mass. 

We restrict the integration of IMF to the mass range $10-20\, M_{\odot}$, since Case~C mass transfer occurs only within this donor mass range in our binary grid as shown in Figure~\ref{mt_class_q09}. We first evaluate the relative number of progenitors in different mass ranges based on the IMF. In particular, the fraction of core collapse supernova progenitors in the mass range $10-20\, M_{\odot}$ is
\begin{equation}
    f_{10\text{-}20/\rm CCSN}
    =
    \frac{N_{10\text{-}20}}{N_{\rm CCSN}}
    =
    \frac{\int_{10\,M_\odot}^{20\,M_\odot} \xi(M)\,dM}
         {\int_{10\,M_\odot}^{40\,M_\odot} \xi(M)\,dM}
    \simeq 0.71 ,
\end{equation}
where we define
\begin{equation}
    N_{\rm CCSN}
    =
    \int_{10\,M_\odot}^{40\,M_\odot} \xi(M)\,dM .
\end{equation}
The expected number of binaries in a given evolutionary channel is written as
\begin{equation}
    \mathcal{N}_{\rm ch} = f_{\rm int} \int \xi(M_1)\, p(a)\,
    \mathcal{I}_{\rm ch}(M_1,a)\,
    dM_1\,da ,
\label{eq:int}    
\end{equation}
where $f_{\rm int}$ is the fraction of massive stars that undergo binary interaction, $p(a)$ is the orbital-separation distribution, and $\mathcal{I}_{\rm ch}$ is an indicator function that takes the value of unity if the model belongs to a given channel and zero otherwise. We adopt $f_{\rm int}=0.7$, motivated by observational studies of massive star binaries \citep{Sana2012}. For orbital separation, we adopt a distribution that is flat in logarithmic separation,
\begin{equation}
    p(a) \propto \frac{1}{a},
\end{equation}
which corresponds to a uniform distribution in $\log a$, i.e., $dN \propto d\log a$. This form, commonly referred to as \"{O}pik's law, is widely adopted as a fiducial first-order approximation for binary populations (e.g., \citet{Poveda2007, Kouwenhoven2010} ). Equation~\ref{eq:int} is discretized and evaluated on our binary grid as
\begin{equation}
    \mathcal{N}_{\rm ch} = f_{\rm int} \sum_i \left[ \int_{M_{1,i,-}}^{M_{1,i,+}} \xi(M_1)\,dM_1 \right]
    \left[ \int_{a_{i,-}}^{a_{i,+}} p(a)\,da \right]
    \mathcal{I}_{{\rm ch},i}.
\end{equation}
Here, the subscript $i$ labels each binary model, while $i,-$ and $i,+$ denote the lower and upper boundaries of the primary mass and orbital separation associated with that model. Applying this procedure to our binary grid of $q=0.9$ with $M_{1}=10-20\,M_\odot$, we find that the fraction of interacting binaries weighted IMF and separation that undergo Case~C mass transfer is
\begin{equation}
    f_{\rm Case\,C|int}
    =
    \frac{\mathcal{N}_{\rm Case\,C}}{\mathcal{N}_{\rm Case\,B}+\mathcal{N}_{\rm Case\,C}}
    \simeq 0.27 .
\end{equation}
Finally, the fraction of CCSN progenitors that experience Case~C mass transfer can be written as
\begin{equation}
    f_{\rm Case\,C/CCSN}
    =
    f_{10\text{-}20/\rm CCSN}
    \times f_{\rm int}
    \times f_{\rm Case\,C|int}.
\end{equation}

Substituting the above values, we obtain
\begin{equation}
    f_{\rm Case\,C/CCSN}
    \simeq
    \underbrace{0.71}_{\text{IMF fraction}}
    \times
    \underbrace{0.7}_{\text{interacting fraction}}
    \times
    \underbrace{0.27}_{\text{Case C fraction}}
    \simeq 0.13 .
\end{equation}
This suggests that Case~C mass transfer may contribute to approximately \(13\%\) of all CCSN progenitors.

We emphasize that this estimate is intended only as a rough assessment based on our models, rather than a substitute for a comprehensive binary population synthesis calculation. Although our grid adopts observational-motivated values of mass ratio and separation distribution for models, a more realistic population synthesis would require sampling the full distributions of orbital period and mass ratio. In addition, the mapping between the mass transfer channel and supernova subtype is not one-to-one. Some Case~B binaries may retain a certain amount of hydrogen envelope, while some Case~C binaries may enter CEE and produce stripped-envelope or Type~IIb supernovae rather than ordinary hydrogen-rich Type~II events (e.g., the scenario suggested for SN 2014C in Section~\ref{sec4.3}). Therefore, we cannot directly associate Case~B with Type~Ib/c or Case~C with Type~II supernovae. Instead, the above estimate should be interpreted as the potential contribution of the Case~C channel to the overall CCSN population. This result suggests that Case~C mass transfer may account for $\sim 13\%$ of all CCSN progenitors, rather than representing a rare channel. 
We note that this fraction is more than an order of magnitude larger than the estimate by \citet{ouchi2017}; they estimated the fraction by a specific binary model grid with the fixed primary ZAMS mass of $16 M_\odot$, but the present work shows that the case C transfer is increasingly more common toward the lower mass that dominates the total SN rate for the standard IMF. 

\subsection{Uncertainties in the Mass Transfer Efficiency}

The adopted mass transfer efficiency primarily determines how the transferred material is partitioned between the companion star and the CSM. Because Case~B and Case~C interactions are defined by the evolutionary state of the donor star at the onset of RLOF, the occurrence rate of Case~C systems is expected to be insensitive to the assumed mass transfer efficiency. Consequently, our estimate that $\sim13\%$ of CCSN progenitors undergo Case~C interaction is unlikely to change significantly under different efficiency prescriptions.

Nevertheless, the mass transfer efficiency can influence the subsequent evolution of the binary system. A higher mass-transfer efficiency allows the companion to accrete a larger fraction of the transferred material, potentially leading to more substantial expansion of the companion star and increasing the possibility of unstable mass transfer or CEE. In contrast, a lower efficiency results in a larger fraction of the transferred mass being lost into the CSM. As a result, the adopted efficiency may affect both the final properties of the supernova progenitor and the structure of the resulting CSM. In reality, the mass-transfer efficiency is expected to depend on the thermal response, rotational evolution, and internal structure of the accreting companion. A self-consistent treatment of these effects would require multidimensional hydrodynamic simulations and remains beyond the scope of this study.

\section{Conclusion}

We have carried out a systematic study of binary stellar evolution using \texttt{MESA} to investigate how binary interaction shapes the circumstellar environment immediately prior to core collapse. Our results identify a clear and robust pathway: Case~C mass transfer, initiated after core helium ignition, naturally produces the dense CSM required for some of the strongly interacting supernovae, including SNe~IIn and rebrightening supernovae.

Across our model grid, binaries with donor masses of $10$--$20\,M_\odot$ and separations of $\sim 1000$--$2700\,R_\odot$ undergo late-stage RLOF within $\sim 10^{3}$--$10^{4}\,$yr before explosion. This interaction ejects $\sim 0.01$--$0.2\,M_\odot$ of material and forms CSM extending to $\sim 10^{16}$--$10^{18}\,$cm, in quantitative agreement with the properties inferred for events such as SN~2014C. Furthermore, our results suggest that Case~C mass transfer may account for $\sim 13\%$ of all CCSN progenitors, providing a viable pathway for producing interacting supernovae. In contrast, earlier mass transfer (Case B) occurs too far in advance of core collapse to influence the immediate pre-supernova environment, while single-star winds are generally insufficient to produce the required densities.

These results demonstrate that late-stage binary interaction operates at the right time, mass scale, and spatial extent to shape the environments of interacting supernovae, without invoking finely tuned eruptive mass-loss mechanisms. In this picture, binary evolution provides a direct and physically motivated link between stellar structure and the dense CSM inferred from observations.

At the same time, our models highlight key uncertainties. The geometry and dynamics of Roche-lobe overflow outflows, including the potential formation of circumbinary structures and the role of radiative cooling, may significantly affect the spatial distribution and confinement of the CSM. Reproducing the most compact CSM configurations inferred in some events likely requires mass transfer occurring even closer to core collapse or a more efficient confinement of the outflow.

Overall, our results establish Case C mass transfer as a key channel for producing dense circumstellar environments and provide a concrete evolutionary framework for interpreting interacting supernovae. Future work combining multidimensional hydrodynamics with binary evolution models, together with systematic observational constraints, will be essential for fully connecting pre-supernova mass loss to the diverse phenomenology of interacting transients.

\begin{acknowledgments}
This research is supported by the National Science and Technology Council, Taiwan, under grant No. MOST 110-2112-M-001-068-MY3, NSTC 113-2112-M-001-028-, 114-2112-M-001-012-, 114-2811-M-001-094, and the Academia Sinica, Taiwan, under a career development award under grant No. AS-CDA-111-M04. ST acknowledges support from Academia Sinica, Taiwan, through the PhD student support program, and from the Japan–Taiwan Exchange Association through the 2025 PhD Student Visiting Program. KC acknowledges the support of the Alexander von Humboldt Foundation. KM acknowledges support from the Japan Society for the Promotion of Science (JSPS) KAKENHI grant Nos. JP24KK0070, JP24H01810, and JP23H04894. 
Our computing resources were supported by the National Energy Research Scientific Computing Center (NERSC), a U.S. Department of Energy Office of Science User Facility operated under Contract No. DE-AC02-05CH11231, and the KAWAS Cluster at the Academia Sinica Institute of Astronomy and Astrophysics (ASIAA). The work of FKR  is supported by the Klaus Tschira Foundation, by the Deutsche Forschungsgemeinschaft (DFG, German Research Foundation) -- RO 3676/7-1, project number 537700965, and by the European Union (ERC, ExCEED, project number 101096243). Views and opinions expressed are, however, those of the authors only and do not necessarily reflect those of the European Union or the European Research Council Executive Agency. Neither the European Union nor the granting authority can be held responsible for them. 
\end{acknowledgments}

\bibliographystyle{yahapj}
\bibliography{refs}

\appendix

\section{Dependence on Mass Ratio}

\begin{figure*}
\centering
\includegraphics[width=1.\textwidth]{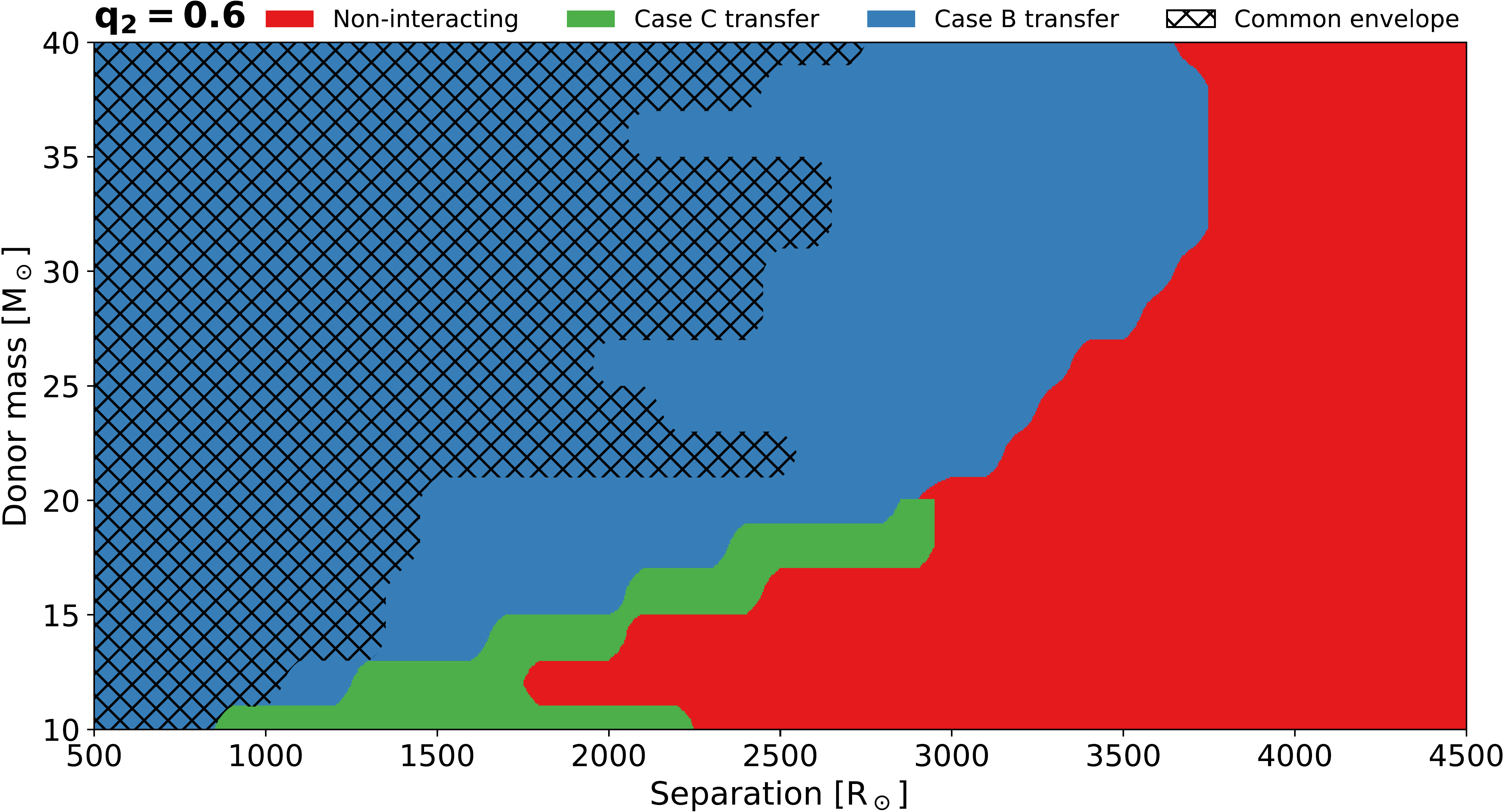}
\label{mt_class_q06}
\caption{Same as Figure~\ref{mt_class_q09}, but for a mass ratio of $q_{2} = 0.6$. Compared to the $q_{2} = 0.9$ case, the overall distribution of mass transfer types remains similar, with Case~C transfer still confined to $M_{1} \lesssim 20~M_{\odot}$. These results indicate that the overall landscape of mass transfer is relatively insensitive to the mass ratio. The main difference is that a smaller $q_{2}$ reduces the overall fraction of interacting binaries while promoting unstable mass transfer and increasing the fraction of CEE.}
\end{figure*}

Figure~\ref{mt_class_q06} shows the distribution of the binary interactions results for a mass ratio of $q_{2} = 0.6$, compared to the case $q_{2} = 0.9$ presented in Figure~\ref{mt_class_q09}.
Among the 656 models, 65.2\% undergo mass transfer, including 59.8\% Case~B and 5.4\% Case~C, which is slightly lower than the corresponding fractions for $q_{2} = 0.9$. This difference can be understood in terms of the Roche-lobe radius of the donor star (Equation~\ref{eq:roche_lobe}). For a fixed orbital separation, a smaller mass ratio $q_{2}$ corresponds to a larger Roche-lobe radius. As a result, binaries with $q_{2} = 0.9$ are more likely to initiate RLOF than those with $q_{2} = 0.6$, leading to a slightly higher fraction of interacting binaries.

The overall distribution of mass transfer types remains qualitatively similar between the $q_{2} = 0.6$ and $q_{2} = 0.9$ grids, with both indicating that Case~C mass transfer is largely confined to donor masses $\lesssim 20~M_{\odot}$. The primary difference lies in the extent of CEE. For $q_{2} = 0.6$, a significantly larger fraction of models enter CEE, particularly among Case~B binaries. This behavior can be understood from the orbital response to mass transfer during RLOF.
Mass transfer is initiated when the donor star expands and fills its Roche lobe. As mass is transferred from the donor to the companion, the orbital separation evolves according to angular momentum conservation. For binaries with an initial mass ratio $q_{2} = M_{2}/M_{1} < 1$, the orbit typically shrinks during the early phase of mass transfer, reaching a minimum when the two stars have comparable masses. Continued mass transfer may subsequently lead to orbital expansion, and the RLOF phase eventually ceases as the donor contracts or loses its hydrogen envelope. For binaries with a smaller initial mass ratio, the initial orbital shrinkage is more obvious. This leads to higher mass transfer rates and increases the probability that the mass transfer becomes dynamically unstable, triggering CEE.

Overall, we find that the mass ratio has only a minor effect on the classification of mass transfer types, but dominates the stability of mass transfer. Binaries with smaller mass ratios tend to undergo common-envelope evolution.

% Figure 5

\end{document}